\documentclass[aps,prb,reprint,showpacs,showkeys,superscriptaddress,preprintnumbers,amsmath,amssymb,longbibliography]{revtex4-2}
\usepackage{graphicx}
\usepackage{dcolumn}
\usepackage{amsmath} 
\usepackage{amssymb}
\usepackage{color}
\usepackage{soul}
\usepackage{xcolor}
\usepackage{multirow}
\usepackage{chemfig}
\usepackage{array}
\usepackage{tabularx}

\usepackage{bm}

\begin{document}

\title{Microscopic Origin of Random Singlet Behavior in B-site Disordered Spin-1/2 Perovskite $\rm BaCu_{1/3}Nb_{2/3}O_3$ Revealed by EXAFS  and Thermodynamics}

\author{Sagar Mahapatra}
\affiliation{Department of Physics, Indian Institute of Science Education and Research, Pune 411008, Maharashtra, India}

\author{Francesco De Angelis}
\affiliation{Dipartimento di Scienze, Universitá Roma Tre, I-00146 Roma, Italy}

%\author{Pramod R Nadig}
%\affiliation{Manipal Institute of Technology, Manipal Academy of Higher Education, Manipal 576104, India}

\author{Martin Etter}
\affiliation{Deutsches Elektronen-Synchrotron (DESY), Notkestra\ss e 85, 22607 Hamburg, Germany}

\author{Edmund Welter}
\affiliation{Deutsches Elektronen-Synchrotron (DESY), Notkestra\ss e 85, 22607 Hamburg, Germany}

\author{M. P. Saravanan}
\affiliation{UGC-DAE Consortium for Scientific Research, University Campus, Khandwa Road, Indore 452 001, India}

\author{Rajeev Rawat}
\affiliation{UGC-DAE Consortium for Scientific Research, University Campus, Khandwa Road, Indore 452 001, India}

\author{Carlo Meneghini}
\affiliation{Dipartimento di Scienze, Universitá Roma Tre, I-00146 Roma, Italy} 

\author{Surjeet Singh}
\email{surjeet.singh@iiserpune.ac.in}
\affiliation{Department of Physics, Indian Institute of Science Education and Research, Pune 411008, Maharashtra, India}
%\affiliation{IFW Dresden, Helmholtzstr. 20, 01069 Dresden, Germany}

\begin{abstract}
\textcolor{black}{
We report a combined structural and thermodynamic study of the ABO$_3$-type disordered perovskite BaCu$_{1/3}$Nb$_{2/3}$O$_3$ (BCNO), whose B site is jointly occupied by Cu and Nb in the $1:2$ ratio. Using synchrotron powder x-ray diffraction (XRD) and x-ray absorption fine structure (XAFS) spectroscopy, we investigate the microscopic nature of Cu$^{2+}$/Nb$^{5+}$ disorder on the pseudo-cubic B-sublattice and its relation to the emergent random-singlet (RS) behavior evidenced at low temperatures. While XRD reveals no long-range Cu/Nb ordering and average site occupancy consistent with stoichiometry, XAFS reveals a peculiar local chemical order characterized by preferential heteroatomic Cu$:$Nb correlations. This local arrangement strongly suppresses direct Cu$:$Cu linkages, despite the Cu concentration being close to the percolation threshold of a cubic lattice. The resulting exchange network explains the absence of spin-glass freezing or long-range magnetic order in the presence of substantial antiferromagnetic interactions, as indicated by a Curie-Weiss temperature $\Theta_{CW}\approx -50$ K. Instead, the magnetic susceptibility $\chi(T)$ and specific heat $c_p(T)$ exhibit power-law behavior and characteristic single-parameter $T/H$ scaling over broad temperature and magnetic-field ranges, consistent with random-singlet phenomenology. Notably, at very low temperatures, the specific heat behavior transitions from $T^{1-\gamma}$ ($\gamma \approx 0.6$ from the $T/H$ scaling) in zero-field to a T-linear dependence under high field, indicating a crossover to a distinct low-energy regime whose microscopic origin remains to be established.} 
\end{abstract}
\maketitle
\section{Introduction} 
Disorder in quantum systems causes exotic ground states, uncharacteristic of the pristine system, to emerge. The Heisenberg antiferromagnetic spin-1/2 chain is a canonical model system where the effects of quenched disorder have been extensively studied. Motivated by early experimental observations of unusual low-temperature power-law behavior in the magnetic susceptibility ($\chi$) and heat capacity ($c_p$) of certain organic spin-chain compounds, Ma, Dasgupta, and Hu introduced the real-space renormalization-group (RG) technique to explain these thermodynamic scaling properties~\cite{ma1979}.\\

In their model, they considered a spin-1/2 chain with quenched disorder, which results in a broad probability distribution $P(J)$ of antiferromagnetic exchange couplings $J$ between the neighboring spins. They showed that the ground state of this system, which is a Tomonaga-Luttinger spin liquid in its pristine form, transits to a random-singlet (RS) phase upon disordering. The RS phase is characterized by the presence of spin-singlets formed over arbitrarily large separations~\cite{Dasgupta_PhysRevB.22.1305,ma1979}.\\ 

Schematically, the real-space RG proceeds by iteratively decimating the strongest exchange $J_0$ in the chain. The two spins connected by $J_0$ form a singlet, generating an effective coupling $\Tilde{J}$ between the spins on either side of $J_0$. Under these iterative transformations, the distribution of renormalized exchange couplings becomes increasingly broad, and the system evolves into a network of spin-singlets. Because the location of the strongest exchange coupling at any given step is random, the resulting singlets form over arbitrarily large separations. At low temperatures, the spins that have not yet been decimated ($J < T$), the so-called surviving spins, dominate the magnetic susceptibility and specific heat response, leading to power-law behaviors characteristic of a  random-singlet ground state~\cite{Dasgupta_PhysRevB.22.1305,ma1979, Fisher1994,doty1992effects,fisher1995critical}.\\ 

Since the presence of disorder is inevitable in any real quantum system, the question of RS-type phase in 3D spin 1/2 systems has remained a topic of significant interest, as reviewed by Vojta et al.~\cite{Vojta2006}. Very recently, Kimchi et al. revisited the specific heat data for a number of geometrically frustrated systems and found that in all cases the heat capacity $c_p(H, T)$ in temperature $T$ and magnetic field $H$ exhibits $T/H$ data collapse reminiscent of scaling near a critical point~\cite{kimchi_nat_com_scaling}. \\

This striking observation led them to propose a theory based on an emergent random-singlet regime in highly-frustrated quantum magnets with quenched disorder~\cite{kimchi_nat_com_scaling, kimchi_2018}. Due to randomness, while the majority of spin-1/2 sites remain dynamic (akin to the spins in a quantum spin liquid), a small fraction nucleates, forming a random network of spin–1/2 moments~\cite{kimchi_nat_com_scaling}. At low energies, the effective coupling between these spins follow a broad power-law probability distribution P($\Tilde{J}$)~$\sim$~$\Tilde{J}^{-\gamma}$, analogous to the spin chain case. Consequently, at very low temperatures, power-law behaviors emerge with $\chi~\propto~T^{-\gamma}$, and $c_p \propto T^{1-\gamma}$, where $\chi$ and $c_{p}$ are the susceptibility and specific heat of the spin system. Furthermore, a single-parameter scaling collapse of the magnetization ($M$) and specific heat ($c_p$) is expected over wide range of temperatures and fields. More precisely, MT$^{\gamma-1}$ versus $\mu_0$H/T plots collapse on to a single curve, and $(\mu_0H)^\gamma c_{p}/T$ versus $T/\mu_0H$ plots behave similarly. Henceforth, we shall label these as $M(H, T)$ and $c[H, T]$ scalings. \\

Some examples of this phenomenology in geometrically frustrated systems include, quantum spin liquid candidate Herbertsmithite (ZnCu$_3$(OH)$_6$Cl$_2$) that has been argued to exhibit the RS phase due to Zn/Cu antisite disorder~\cite{kimchi_nat_com_scaling}. In the triangular lattice compound, Y\(_2\)CuTiO\(_6\), similar observations of random-singlets are found due to the Cu/Ti disorder~\cite{18_SKundu_YCTO}. YbMgGaO\(_4\) is also claimed to fit in an RS phase-like scenario due to the randomness in the exchange interactions arising from the antisite disorder between Mg and Ga ions, affecting randomly the exchange couplings between the Yb moments on a frustrated triangular lattice ~\cite{YbMg_2017}. A similar bond exchange randomness induced by Li/Zn antisite disorder stabilizes an RS-like phase in LiZn$_2$Mo$_3$O$_8$~\cite{LiZn2012}.\\

This work is motivated by recent studies on Cu$^{2+}$ (spin-1/2) based disordered perovskites SrCu$_{1/3}$M$_{2/3}$O$_3$ (M = Ta, Nb). Hereafter, referred to as SCTO and SCNO, respectively. The crystal structure of these ABO$_3$-type perovskites is tetragonal, featuring a nearly cubic ($c/a\sim 1$) B-sublattice %, randomly occupied by 
with Cu and Nb (or Ta) occupancy in the $1:2$ ratio. Both these compounds are shown to exhibit characteristics of a random-singlet ground state~\cite{hossain2024evidence,sana2024possible}. This is notable, because unlike the two-dimensional frustrated triangular lattice encountered in the previous examples, the RS phase here is realized on a three-dimensional, unfrustrated network. Furthermore, at 33\%, the concentration of magnetic Cu atoms in these compounds exceeds the percolation threshold (31\%) of the cubic-lattice. One thus expects the presence of large, ideally infinite, Cu-connected networks, leading to long-range magnetic ordering or spin-glass-like freezing of the Cu-spins, rather than the experimentally observed RS phase.\\   

In this paper, we present a detailed structural and thermodynamic study of BaCu$_{1/3}$Nb$_{2/3}$O$_3$ (henceforth BCNO), a compound structurally related to its Sr-based analogues. Synchrotron x-ray diffraction (XRD) reveals no long-range Cu/Nb ordering and an average B-site occupancy consistent with the nominal 1$:$2 stoichiometry. In contrast, XAFS reveal local details (bond distances, chemical ordering, etc.) essential for developing reliable theoretical models to understand the physical properties of complex quantum materials with quenched disorder. In BCNO, Cu and Nb K-edge XAFS reveals the local chemical arrangement, characterized by preferential Cu$:$Nb heteratomic correlations. This local arrangement strongly suppresses direct Cu$:$Cu connectivity and Cu clustering, despite the relatively high Cu concentration, which is close to the site-percolation threshold of a simple cubic lattice. The specific heat indicates no signs of long-range magnetic ordering down to 0.1 K. Instead, the low-temperature behavior is found to be highly reminiscent of a random-singlet phase with the zero-field specific heat following a T$^{1-\gamma}$ ($\gamma~\sim~ 0.6$) dependence over more than three decades in temperature, and magnetic susceptibility, $\chi(T)$, showing a $T^{-0.6}$ divergence. The RS phase is further corroborated by the $c_p[H, T]$ and $M[H, T]$ scalings. Furthermore, we show that under a modest applied field of 3 T, the specific heat behavior evolves from $T^{0.42}$ ($\mu_0H= 0$) to $T^{1.6}$ ($\mu_0H= 3$~T). As the field strength increases further, the exponent slowly decreases, evolving towards 1 as $\mu_0H \rightarrow 9$~T. The magnetic entropy saturates around 25\% of $R\ln2$, a value that is essentially independent of the applied magnetic field. These observations are interpreted in terms of the presence of two subsystems, as envisaged in Ref.~\cite{kimchi_2018}, with 25\% of the spins forming a random singlet phase
and remaining spins staying in a correlated quantum paramagnetic phase.\\ 
 
 Our study clarifies the microscopic origin of exchange randomness in these AM$_{1/3}$M$'_{2/3}$O$_3$ type three-dimensional disordered perovskites and its role in stabilizing random-singlet behavior.
 \textcolor{black}{Furthermore, the application of high magnetic field seems to reveal the hidden spin liquid-like quantum paramagnetic state, inferred from the linear-in-temperature dependence of the specific heat of the spin system.}
 \\ 

\section{Experimental methods}
\begin{figure*}
    \centering
   \includegraphics[width=0.95\linewidth]{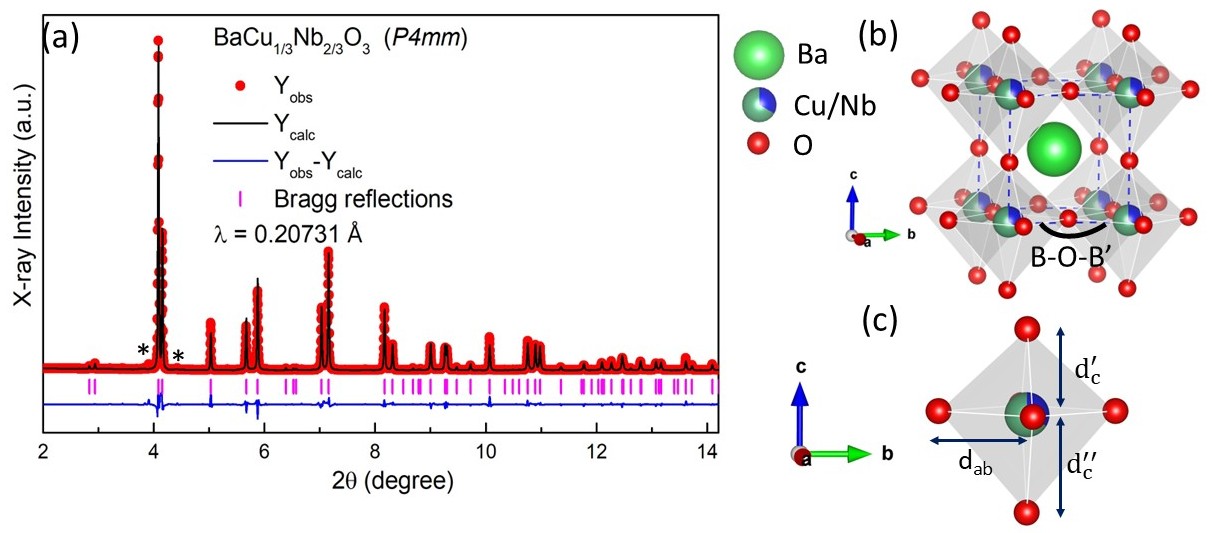}
    \caption{(a) Synchrotron X-ray powder diffraction pattern of BCNO showing observed (red), calculated (black), and difference (blue) plots. The magenta sticks show the position of the expected Bragg peaks. (b) The ABO$_3$-type crystal structure of BCNO. (c) An isolated BO$_6$ octahedron showing the slight off-centering of the B-site cations. }
    \label{fig:xrd}
\end{figure*}

A polycrystalline sample of $\rm BaCu_{1/3}Nb_{2/3}O_3$  was prepared by the standard solid-state reaction method. For this purpose, stoichiometric quantities of BaCO$_3$, CuO, and Nb$_2$O$_5$ powders were thoroughly homogenized using a mortar and pestle, and sintered multiple times at $1100^\circ$C~with intermediate grindings and pelletizing. BaCO$_3$ was preheated at $700^\circ$C and was weighed right after it was taken out of the furnace at $200^\circ$C. The preliminary phase check was done using a Bruker D8 Advance powder diffractometer.\\

A more detailed structural characterization was done using synchrotron based powder X-ray diffraction (XRD) and X-ray absorption spectroscopy (XAS) at P02.1 and P65 beam lines, respectively, at the PETRA III facility of DESY, Hamburg, Germany. For the  XAS measurements, samples were prepared by mixing BCNO fine powders with boron nitride (BN) and pressing into thin solid pellets suitable for handling. Absorption spectra were measured at Cu-K (8979~eV) and Nb-K (18986~eV) edges in transmission geometry using an ionization chamber to measure incident (I$_o$) and transmitted  (I$_1$) x-ray intensities, the absorption signal was calculated as $\mu t$ = ln(I$_0$/I$_1$). Data measurements were carried out four times at each edge under identical conditions to improve reproducibility and enhance the averaging of data statistics. \\

The energy calibration and reproducibility of the measurements were monitored using Cu and Nb metal foils measured simultaneously with the sample. The raw XAS data were processed following the standard procedures for pre-edge subtraction, normalization, background removal, glitch rejection and spectral averaging to obtain the final (X-ray absorption fine-structure spectroscopy) XAFS signal~\cite{meneghini2012estra, EstraPy}. Quantitative analysis of the extended fine structure (EXAFS) region was carried out as described in sec.~\ref{sec:XAS}.\\

The specific heat ($T > 0.1$ K), and magnetization ($T > 3$ K) measurements were carried out up to 9 T of applied magnetic field using a Physical property measurement system (PPMS) Quantum Design, USA. For the specific heat, a run was done down to 2 K in the heat-capacity option; whereas, the dilution attachment was used to extend these measurement down to 0.1 K. Using the dilution unit, the measurements were performed from 0.1 K to 3.5 K, with the data from the two measurements (with dilution attachment and without it) overlapped nicely over the common temperature range. For both measurements, the addenda measurement with Apizon N grease, was taken separately prior to the actual run with the sample. the thermal coupling between the sample and the platform (or sample stage) is shown. A thermal coupling between the sample and the sample platform exceeded 90\% over the whole temperature range, suggesting an excellent thermal coupling throughout the temperature range. The sample and sample-platform temperature variations as a function of time exhibited excellent fits to the two-tau method used in the PPMS to extract the specific heat from the raw data.\\

\section{Results and discussion}
\subsection{Structural characterization} 

\subsubsection{Synchrotron X-ray powder diffraction}
The Rietveld analysis of synchrotron radiation ($\rm \lambda$ = 0.20731~\AA) based powder XRD pattern (Fig.~\ref{fig:xrd}) suggested that our BCNO sample belongs to the perovskite family with the general formula ABO$_3$ with the B-site shared by Cu and Nb in the ratio $1:2$. In literature, several space groups under the tetragonal symmetry have been used for this compound. For example, in Ref~\cite{zhang2006ferroelectric}, the lab-based XRD pattern of BCNO was fitted using five plausible tetragonal space groups, namely: $P422$ (space group no. 89), $P4mm$ (space group no. 99),$P\Bar{\textit{4}}$2m (space group no. 111), $P\Bar{4}m2$ (space group no. 115), and $P4/mmm$ (space group no. 123), with small differences in their R$_{wp}$-values. We found that the high-quality synchrotron data are best described using the non-centrosymmetric space group $P4mm$%(R$_{wp}$ = 9.33\%)
. The structure obtained from the XRD refinement is shown in Fig.~\ref{fig:xrd}(b). The Cu/Nb ions occupy the same crystallographic site in an octahedral environment that lacks inversion symmetry. %due to the off-centering. For clarity, 
An individual (Cu/Nb)O$_6$ octahedron is shown in Fig.~\ref{fig:xrd}(c). The average B-B distance within the $ab$-plane is designated as $d_{ab}$% and that along the $c$-axis as $d_c$
. %In the non-centrosymmetric  \textit{P4mm} structure, t
The average distances of the apical oxygen atoms from the central B atom are designated as $d'_c$ and $d''_c$ ($\neq d'_c$). The atomic positions, occupancies, lattice parameters, bond distances/angles and goodness of fit parameters of the Rietveld refinement are given in the Table~\ref{tab:BCNO_p4mm_REF} and~\ref{Table: XRD_refinement}, respectively. The $c/a$ ratio is marginally greater than 1 ($\approx$ 1.036) and B-O-B bond angle (175$^\circ$) marginally shorter than 180$^\circ$.

Since powder XRD is sensitive only to the long-range periodic structure, the refined atomic positions represent crystallographic averages over the entire sample. Consequently, the local atomic arrangements and short-range correlations that break translational periodicity are lost in the diffraction experiment, mandating the use of a local probe such as XAFS.  It is noteworthy that the concentration of the magnetic Cu$^{2+}$ is 33\%, which exceeds the site-percolation threshold for a cubic lattice. Under such conditions, one expects the formation of extended magnetic networks that could support long-range magnetic order or spin-glass-like freezing in the least. Hence, it is crucial to determine how the Cu ions are distributed within the lattice, i.e., whether they form connected clusters/chains or remain spatially dispersed due to local chemical correlations. The local atomic structure sensitivity and elemental selectivity of XAS make it especially suited to address this question, as discussed in detail in sec.~\ref{sec:XAS}.

\begin{table}
%\caption{}\\
\centering
\caption{Atomic positions, thermal isotropic factors, and occupancies for BCNO at 300 K as determined from
Rietveld refinement of synchrotron powder XRD data using the
tetragonal space group $P4mm$. }
\label{tab:BCNO_p4mm_REF}
\renewcommand{\arraystretch}{1.5}
\begin{ruledtabular}
\begin{tabular}{lllll}
%\begin{tabularx}{\columnwidth}{|l|l|l|l|l|}
Atom & x   & y   & z   & Occupancy \\ \hline

Ba   & 0.5   & 0.5   & 0.519 (1)  & 1         \\
Cu   & 0.0   & 0.0   & 0.014 (2)  & 0.333         \\
Nb   & 0.0   & 0.0   & 0.014 (2)  & 0.667         \\
O (1) & 0.0   & 0.0  & 1.480 (4)  & 1         \\
O (2) & 0.5   & 0.0  & 0.033 (6)  & 1\\
\end{tabular}
\end{ruledtabular}
\end{table}

\begin{table}
\center
\renewcommand{\arraystretch}{1.75}
\caption{The structural parameter obtained using the Rietveld refinement for BaCu$_{1/3}$Nb$_{2/3}$O$_3$.}
\label{Table: XRD_refinement}
\begin{tabular}{ll}\\\hline\hline
\textbf{Compound }& \hspace{2 cm} \textbf{BCNO}  \\\hline

Space group & \hspace{2 cm}$P4mm $ \\\hline

$a$ (\AA)   & \hspace{2 cm}  4.04386 (3)  \\\hline

$c$ (\AA)   & \hspace{2 cm} 4.19104 (5)   \\\hline

$c/a^1$    & \hspace{2 cm}1.036   \\\hline

$V$ (\AA$^3$)   &\hspace{2 cm} 68.535 (1) \\\hline

$d'_c$ & \hspace{2 cm} 1.95587 (4) \\\hline
$d''_c$ &\hspace{2 cm} 2.23517 (5)  \\\hline

$d_{ab}$   & \hspace{2 cm} 2.02365 (4)  \\\hline

B-O-B($^\circ$)    &\hspace{2 cm} 175.2712\\\hline

$\rm \chi^2$     & \hspace{2 cm} 2.75 \\\hline

$\rm R_P$       & \hspace{2 cm} 12.5 \\\hline

$\rm R_{WP}$     &\hspace{2 cm} 14.9  \\\hline\hline                     
                               
\end{tabular}
\end{table}

\subsubsection{Local structure analysis}
\label{sec:XAS}

\begin{figure*}[htp]
\centering
\includegraphics[width=\textwidth]{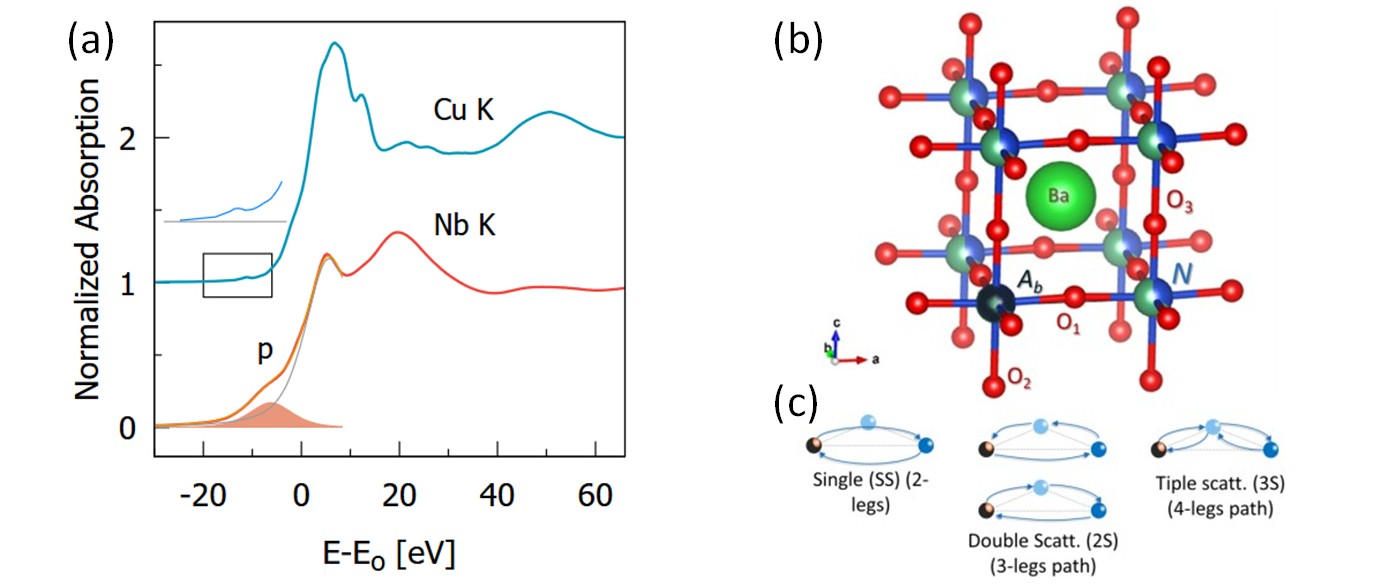}         
    \caption{(a) Normalized XANES spectra measured on BCNO at the Cu-K and Nb-K edges. For clarity, the energy scale is referenced to the corresponding metal foil edge energies, and the spectra are vertically offset. A tiny pre-edge shoulder is visible in the Cu K-edge XANES spectrum, highlighted in the upper inset. A pre-edge shoulder (denoted p) is also observed at the Nb-K edge. The contribution of the pre-edge peak used in fitting the Nb edge features is shown in orange for clarity, together with the model curve for the edge rise (gray line) (see text). (b) The schematic representation of the pseudo-cubic unit of \textit{P4mm} structure with the absorber $A_b$ (Cu/Nb) in the left bottom corner (black) coordinated to oxygen atoms (red). The Ba ions are located at the cube center (green) and 6 neighbor metal N (Cu/Nb) are located on the other cube corners (green/blue). (c) The main single and multiple (double 2S, and triple 3S) photoelectron scattering paths associated to the $A_b$-O-$N$ atomic configurations are presented in the bottom (see text).}
    \label{fig:XANES_model}
\end{figure*}

The X-ray absorption spectroscopy (XAS)
complements XRD by offering chemical selectivity and sensitivity to the local atomic structure~\cite{Rehr_2000}. These local details, in particular metric (bond distances) and chemical order (relative arrangement of specific ions), are essential for developing reliable theoretical models to understand the physical properties of complex quantum materials with quenched disorder. Discussing the  Cu and Nb K edge XAFS data analysis, let us start with the near-edge (XANES) region  that provides detailed information about the electronic state and coordination geometry around the absorber~\cite{SRbook}.\\

%The analysis of XRD patterns describes the periodic structural features of a sample, i.e., its long-range order, but XRD is inherently insensitive to non periodic features and thus it is blind to the so-called local order. However, local order details, both metric (bond distances) and chemical (relative arrangement of specific ions), can significantly influence the macroscopic properties of materials. In particular, elucidating the nature of local order is essential for building reliable models for complex materials, as in the case of disordered quantum systems. The X-ray absorption spectroscopy (XAS) complements XRD by offering chemical selectivity and sensitivity to the local atomic structure.~\cite{SRbook} Analysis of the X-ray absorption fine structure (XAFS) signal, including both the near-edge (XANES, X-ray Absorption Near Edge Structure) and extended (EXAFS, Extended X-ray Absorption Fine Structure) regions, provides accurate information about the oxidation state, coordination geometry, and local structural distortions around specific absorbing atoms. \\ 

\par The normalized XANES spectra of Cu-K and Nb-K edge measured on BCNO sample are presented in Fig.~\ref{fig:XANES_model}(a). The position and shape of the Cu-K edge spectra are consistent with a Cu$^{2+}$ oxidation state and octahedral coordination~\cite{kau1987x}. In particular the pre-edge peak observed in the Cu-K edge XANES (highlighted in Fig.~\ref{fig:XANES_model}a)  represent the dipole forbidden 1\textit{s} - 3\textit{d} transitions and is characteristic of Cu$^{2+}$ while it is absent for Cu$^{1+}$. The shape and position of the Nb-K edge are consistent with Nb$^{5+}$ ions in an octahedral coordination~\cite{shinyoshi2023radiation}. According to Ref.~\cite{XANES_NbTa}, the Nb pre-edge shoulder (labeled p in Fig.~\ref{fig:XANES_model}a) originates from 1$s$ photoelectron excitations into hybridized \textit{pd} states. Although \textit{pd} hybridization is symmetry-forbidden in a perfectly centrosymmetric \textit{$O_h$} environment, it can be enhanced by local structural distortions. The authors of Ref.~\cite{XANES_NbTa} report a negative correlation between the area of the pre-edge peak and the crystal-field splitting of the $e_g$–$t_{2g}$ levels in Nb; that is, while \textit{$O_h$} symmetry reduces the pre-edge peak area, it increases the crystal-field splitting. The Nb edge was modeled with two pseudo-Voigt peaks (with identical mixing factor = 0.5 and FWHM) to account for the pre-edge shoulder p and the first XANES peak, together with a sigmoid step to represent transitions into the continuum. We found the p-peak area (Fig.~\ref{fig:XANES_model} a)  in between 2-2.5 eV suggesting a weak deviations from ideal $O_h$ symmetry and $e_g$–$t_{2g}$ splitting around 3~eV \cite{XANES_NbTa}.

The quantitative analysis of the EXAFS structural signals, $\chi^{\text{exp}}(k)$, was done using the model $\chi^{model}(k) = \sum\limits_{i} \chi^{\text{th}}_i(k)$ which is a linear combination of the scattering contributions  \(\chi^{\text{th}}_i(k)\) (single scattering or SS, and multiple scattering or MS, as shown in Fig.~\ref{fig:XANES_model}c) from various coordination shells~\cite{Rehr_2000}. The SS and MS contributions were calculated using the standard EXAFS equation under the assumption of small Gaussian disorder:
\begin{equation*}\label{eq1}
   \chi^{\text{th}}_i(k) = \frac{S_0^2 N_i}{k R_i^2} f_i(k) \sin[2kR_i + \delta_i(k)]~  e^{-2k^2\sigma_i^2}~e^{-\frac{2R_i}{\lambda(k)}},
\end{equation*}

\noindent where $N_i$, $R_i$, and $\sigma_i^2$ are structural parameters describing the paths geometry. For single scattering paths, $N_i$ denotes the coordination number, $R_i$ the average interatomic distance, and $\sigma_i^2$ the mean square relative displacement (MSRD) that accounts for both thermal and static disorder. For multiple scattering paths, $N_i$ indicates the path multiplicity, $R_i$ is the average half-path length, and $\sigma_i^2$ is the variance of the path length distribution. The scattering amplitude $f_i(k)$,  total phase shift and photoelectron mean free path $\lambda(k)$, are energy-dependent functions originating from the scattering process that were calculated using the FEFF10 program using Hedin-Lundqvist self consistent exchange and correlation atomic potentials~\cite{feff_ankudinov1998real}. The $S_0^2$ is an empirical amplitude reduction factor accounting for many-body losses in the one-electron approximation. The photoelectron wavevector $k$ [$k=\hbar^{-1}\sqrt{m_e (E-E_o+\Delta E}$), where $m_e$ is the electron mass, and $E_o$ the edge-energy], is adjusted by refining the energy shift parameter $\Delta E$, taking into account the mismatch between the experimental edge energy and the Fermi level of theoretical calculation. 

\begin{figure}
    \centering
    \includegraphics[width=\linewidth]{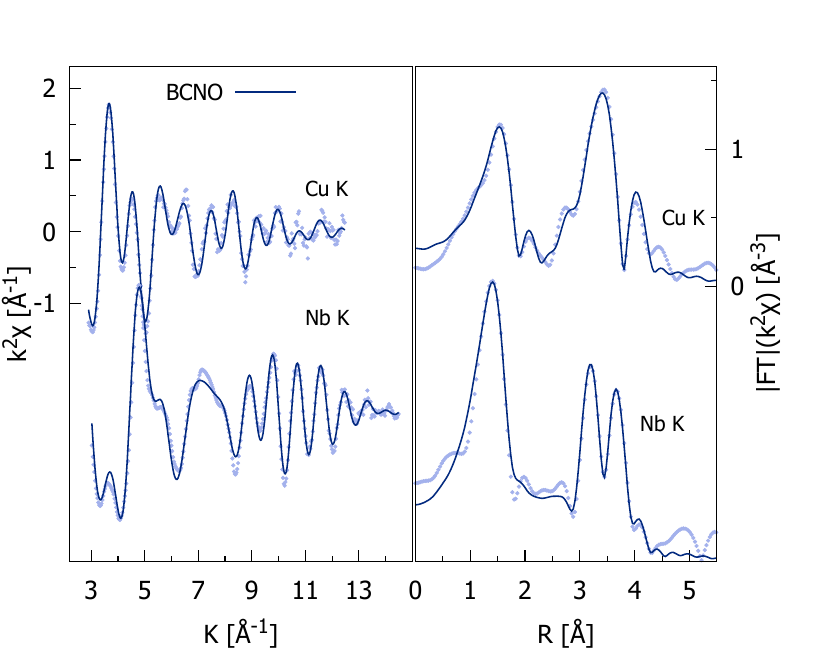}
    \caption{Left panel: \(k^2\)-weighted experimental EXAFS data (dots) and best fit curves (full lines) measured at the Cu and Nb K edges, on BCNO sample. Right panel: moduli of the \(k^2\)-weighted Fourier transforms of experimental data (dots) and best fit curves (full lines). Data are vertically shifted for clarity. }  
    \label{fig:EXAFS_fit}
\end{figure}

The local atomic cluster, employed to compute the theoretical scattering functions and identify the main contributions to the EXAFS model, was built using the $P4mm$ crystallographic structure derived from the Rietveld refinement of the SR-XRD patterns. The EXAFS data were fitted by minimizing the squared residual function  
\(
F = \sum_k \left[ k^2 \left( \chi^{\text{exp}}(k) - \chi^{\text{model}}(k) \right) \right]^2\).  
Fig.~\ref{fig:EXAFS_fit} shows the k$^2$-weighted experimental spectra together with the corresponding best-fit curves, as well as the modulus of their Fourier transforms ($\vert \rm{FT}\vert$). The main peaks around 1.6 \AA\ in Fig.~\ref{fig:EXAFS_fit} are ascribed to the oxygen coordination shell (without phase-shift correction, which apparently downshifts the peak position by about 0.4–0.5 \AA). In this case, the pronounced asymmetry of the Cu–O shell is linked to the strong Jahn–Teller distortion of the Cu$^{2+}$O$_6$ octahedron, while the symmetric Nb–O peak is consistent with the non–Jahn–Teller-active nature of Nb ions. The more intricate structures between 2.5 \AA\ and 4.5 \AA\ arise from several contributions associated with single-scattering (Ba, Nb/Cu and O next neighbors) and multiple-scattering paths, which are enhanced by the focusing effect due to the nearly collinear A$_b$-O-N arrangement along the perovskite cube edge, as illustrated in Fig.~\ref{fig:XANES_model}(b)  

The EXAFS data refinement of BCNO was performed in the same manner as for its Ta-analogue, $\rm BaCu_{1/3}Ta_{2/3}O_3$, using structural constraints to minimize parameter correlations and to achieve stable, reliable information on the local coordination environments of Cu and Nb. A detailed description of the analysis procedure can be found in the supplementary information of Ref.~\cite{mahapatra2026emergent} for readers seeking further information. In what follows, we concentrate on the results that are directly relevant for understanding the physical properties of BCNO. The EXAFS analysis indicates that the local surroundings of Cu and Nb differ from the average periodic structure of the unit cell. This is expected, since Cu$^{2+}$ and Nb$^{5+}$ differ significantly in their valence state, ionic radius, and Jahn-Teller activity. In particular, we found four shorter Cu-O$_1$ bonds ($\rm R_{Cu-O_1} \approx$ 2.03~\AA) and two longer Cu-O$_2$ bonds ($\rm R_{Cu-O_2} \approx$ 2.30~\AA) in CuO$_6$ octahedra, in agreement with the strong Jahn-Teller active nature of the Cu$^{2+}$ ion. In contrast, the NbO$_{6}$ octahedron appears much less distorted with 4 $O_1$ at $\rm R_{Nb-O_1} \approx$ 1.92~\AA and 2 $O_2$ at $ \rm R_{Nb-O_2} \approx$ 1.90~\AA, consistent with the non JT active nature of Nb$^{5+}$. The comparatively longer Cu-O bond length relative to Nb is consistent with the larger ionic radius of Cu$^{2+}$ (0.73~\AA) compared with Nb$^{5+}$ (0.64~\AA) in an octahedral coordination. The Cu/Nb–Ba distances are found to be 3.53~\AA, in agreement with the XRD results, while a more distant O$_3$ shell at about 4.5~\AA\ is required in the fit, arising from the 24 oxygen atoms located on the other edges of the cube. Both single (SS) and multiple (MS) scattering contributions from A$_b$-O-N configurations ($A_B$ designates absorber which is either Cu or Nb, and N = Cu/Nb, see Fig.~\ref{fig:EXAFS_fit}b), are included in the analysis to extract structural %(bond lengths and angles) 
and chemical %(relative neighbor probability) 
information listed in Table.~\ref{table:XAFS}. %The results indicate that the Cu/Nb-O-Cu/Nb bond angle, [$\Theta_{Cu/Nb-O-Cu/Nb} =2\sin^{-1}(\frac{R_{SS}}{R_{3S}})$], departs from the ideal 180$^{\circ}$ and is approximately 157$^{\circ}$, implying a slight tilt of the octahedra.
\\ 

More relevant is the question of chemical ordering, namely how Cu and Nb are distributed on the lattice. A comprehensive treatment of the chemical order analysis is given in Ref.~\cite{mahapatra2026emergent}. Here, we note that the EXAFS results demonstrate that the Cu and Nb are arranged so as to maximize the number of hetero-atomic scattering paths. In fact, around Cu, we found that more than 90\% of the nearest neighbor sites are occupied by Nb, with $N_{\text{Cu}-\text{Nb}} > 5.4$ (and $N_{\text{Cu}-\text{Cu}} < 0.6$). On the other hand, around Nb, we found an equal number of Nb and Cu neighbors: $N_{\text{Nb}-\text{Cu/Nb}} = 3$. This finding is essential for understanding the physical properties of these complex quantum systems. \\

Indeed, although the Cu concentration lies above the percolation threshold, which would imply the possibility of large Cu clusters, the condition $N_{\text{Cu}-\text{Cu}} < 0.6$ suggests that the probability of finding a Cu adjacent to another Cu is low ($<$ 10\%). Hence the number density of Cu::Cu dimers (omitting the intervening O atom for brevity) is expected to be small, while Cu::Cu::Cu trimers and larger Cu clusters become progressively rarer. On the other hand, the nearly equal probability of finding Cu or Nb around an Nb site ($N_{\text{Nb}-\text{Cu/Nb}} = 3$) favors the likelihood of forming Cu::Nb::Cu units, while configurations with two or more intervening Nb atoms are expected to be less frequent than in a completely random distribution. \\

In the language of percolation theory, such connectivities, as encountered here, result in a large concentration of monomers as only a small fraction of Cu ions are directly connected to another Cu ion in their nearest neighborhood. Taking $N_{Cu-Cu} < 0.6$, the probability that a Cu finds another Cu at a nearest-neighbor site around it is $\rm p_{Cu-Cu}$ is 0.1. Assuming uncorrelated local occupancy, the probability that a Cu ion has no Cu nearest neighbors is $\rm p_{monomer} = (1 - p_{Cu-Cu})^6 = 0.9^6 = 0.53$, implying a monomer fraction exceeding 50\%. Given that the occupancy is not random but exhibits strong heteroatomic correlations, this estimate should be regarded as a lower bound. Thus, the structure essentially consists of a large number of Cu monomers, together with a much smaller population of dimers, an even smaller number of trimers, and so on. This picture is in excellent agreement with Ref.~\cite{hossain2024evidence}, where model calculations indicated that a monomer concentration of approximately 63\% is required to reproduce the low-temperature specific heat of SCNO consistent with its random-singlet ground state.

\begin{table}[htb]
\caption{Results of the EXAFS data refinement at the Cu-K and Nb-K edges. Local structure parameters are compared with crystallographic distances based on P$4mm$ crystallographic structure. Standard uncertainties on the last digits of refined parameters are shown in parentheses.}
\label{table:XAFS}
    \centering
%\fontsize{5}    
\renewcommand{\arraystretch}{1.2}
\begin{ruledtabular}
    \begin{tabular}{lllll}
       shell & P$4mm$      & \multicolumn{3}{l}{ {Cu K-Edge}} \\
        \hline
       & R\textsubscript{XRD}      & N & R     & \(\sigma^2\)  \\
       & (\AA)  &   & (\AA) & \(\times 10^2\) (\AA\textsuperscript{2}) \\
        \hline
O$_1$  & 2.022       &4      & 2.03(1)    & 0.91(4)   \\
O$_2$  & 2.095       &2      & 2.30(2)    & 0.91      \\
Ba     & 3.50        &8      & 3.543(8)   & 1.6(3)    \\ 
Nb(SS) & 4.04/4.19   &5.4(3) & 4.04(2)    & 2.9(2)    \\
Nb(2S) &             &10.8   & 4.14       & 2.0       \\
Nb(3S) &             &5.4    & 4.24       & 1.1(1)    \\
Cu(SS) & 4.04/4.19   &0.6    & 4.15(1)    & 0.40(3)   \\
Cu(2S) &             &1.2    & 4.19       & 0.69      \\
Cu(3S) &             &0.6    & 4.23(2)    & 0.98(8)   \\
O$_3$  & 4.55        &24     & 4.43(4)    & 2.0(4)    \\
        \hline    
        &       & \multicolumn{3}{l}{ {Nb K-Edge}} \\
        \hline
       & R\textsubscript{XRD}      & N & R     & \(\sigma^2\)  \\
       & (\AA)  &   & (\AA) & \(\times 10^2\) (\AA\textsuperscript{2})  \\
        \hline
O$_1$     & 2.022     &4     & 1.918(5)   & 0.89(3)   \\
O$_2$     &           &2     & 1.901(5)   & 0.89      \\ 
Ba        & 3.50      &8     & 3.528(8)   & 0.91(2)   \\  
Nb(SS) & 4.04/4.19    &3     & 3.92(2)    & 0.45(3)   \\
Nb(2S) &              &6     & 3.97       & 2.3       \\
Nb(3S) &              &3     & 4.01       & 4.0(2)    \\
Cu(SS) & 4.04/4.19    &3     & 4.00(2)    & 1.4(3)    \\
Cu(2S) &              &6     & 4.10(1)    & 0.8       \\
Cu(3S) &              &3     & 4.20       & 0.4(1)    \\
O$_3$  & 4.50         &24    & 4.50(2)    & 2.2(2)    \\
         
\end{tabular}
\end{ruledtabular}
\end{table}

\subsection{Magnetic and thermodynamic behavior} 
\subsubsection{Magnetization} 
\begin{figure*}
   \centering
   \includegraphics[width= \linewidth]{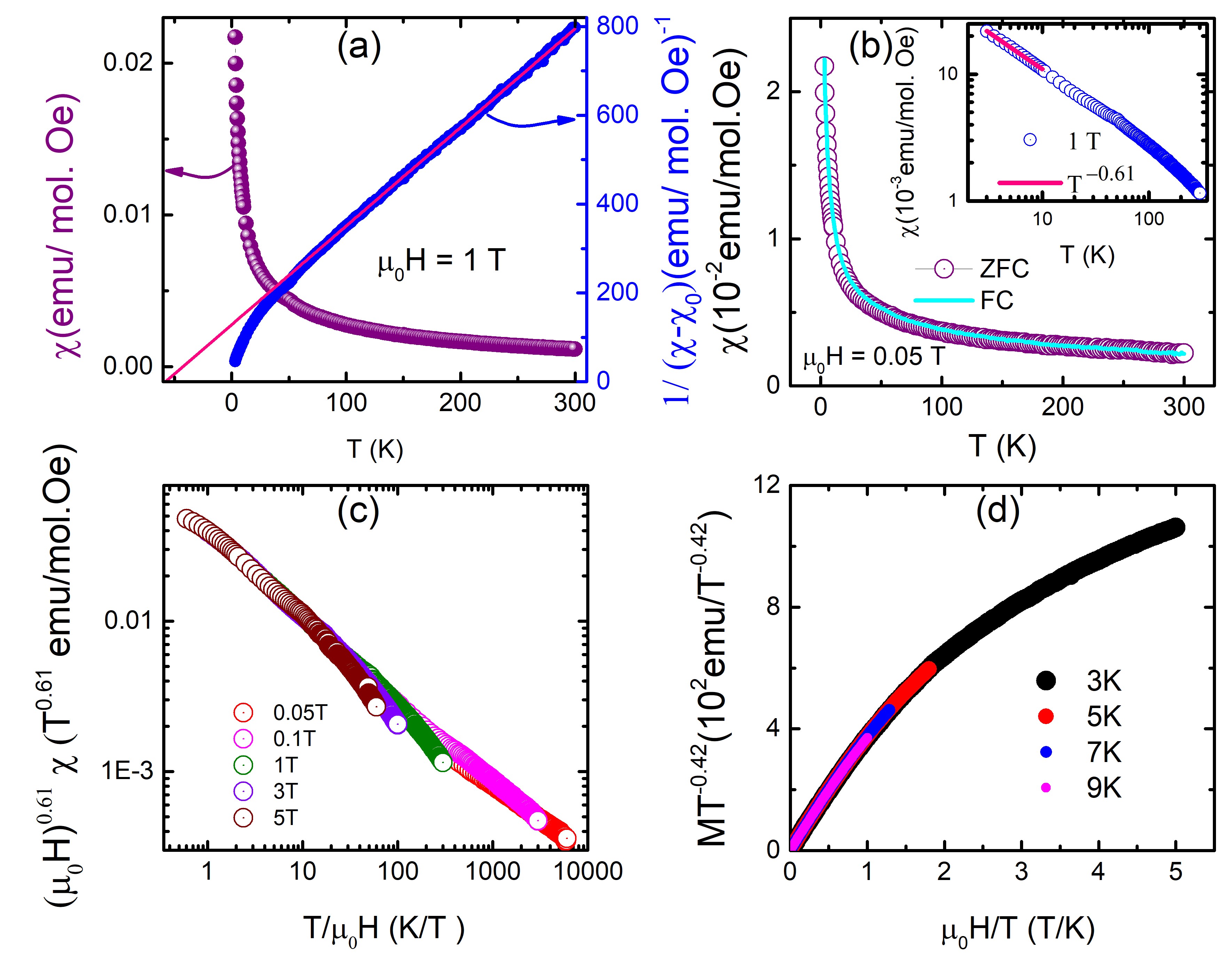}
  \caption{Magnetic susceptibility and T/H scaling in BCNO. (a) The temperature variation of magnetic susceptibility $\chi(T)$ and the inverse magnetic susceptibility $1/(\chi - \chi_0)$. The pink line through the data points is the Curie-Weiss fit. (b) The zero-field-cooled (ZFC) and field-cooled (FC) data measured under $\mu_0H = 0.05$T. Inset: power-law ($T^{-\gamma}$) fitting of $\chi-\chi_0$ below 10~K shown a log-log scale. (c) and (d) show data scaling behavior of $\chi(T)$ and isothermal magnetization $M(H)$, respectively. See text for details.} 
  \label{fig:chi_BCNO}
\end{figure*}

The temperature dependence of magnetic susceptibility ($\chi$) of BCNO, measured under an applied magnetic field of 1~T in the temperature range 3 $\le$ T $\le$ 300~K, is shown in  Fig.~\ref{fig:chi_BCNO}(a). No sign of long-range ordering or spin-glass-like freezing could be observed in the measured temperature range. The $\chi$(T) data in the high temperature region above 100 K is fitted using the modified Curie-Weiss expression:      
\begin{equation} \label{Equation: Eq1}
 \rm \chi(T) =
 \chi_0 + \frac{C}{T-\Theta_{CW}},
\end{equation}
 where $\chi_0$ represents the temperature independent contribution to the susceptibility due to core-diamagnetism and van Vleck-type paramagnetism, $C$ is the Curie constant and $\rm \Theta_{CW}$ is the characteristic Curie-Weiss temperature. The red line through the data points overlaying the inverse $\chi -\chi_0$ plot depicts the Curie-Weiss behavior, which is extrapolated to lower temperatures to highlight the departure from the Curie-Weiss behavior. The zero-field-cooled (ZFC) and field-cooled (FC) plots overlap over the whole temperature range, substantiating the absence of any spin-glass-like phase due to structural randomness. This observation is also consistent with the absence of any magnetic clusters, as inferred from our detailed EXAFS analysis. The best-fit values of Curie-Weiss parameters are: $\chi_0 = -6.5 \pm 0.5~\times~10^{-5}$~emu~mol$^{-1}$ Oe$^{-1}$, $\rm \theta_{CW} = -50~\pm~10$ K, and $\rm C = 0.44~\pm~0.03$~emu~mol$^{-1}$ Oe$^{-1}$ K, from which the effective magnetic moment ($\rm \mu_{eff}$) of Cu is calculated to be $\rm 1.87 \pm 0.06~\mu_B$. The negative Curie-Weiss temperature indicate the dominant antiferromagnetic correlations, in line with the empirical Goodenough-Kanamori rule concerning the sign of exchange correlation vis-a-vis the B-O-B bond angle in the $ab$-plane ($\sim$175$^{\circ}$) and along the c-axis $180^{\circ}$ ~\cite{goodenough1958interpretation}.\\
 
 The magnitude of $\rm \Theta_{CW}\sim50$~K, suggests a moderate to strong magnetic correlation among the spins, yet no sign of long-range ordering down to 0.1~K (as inferred from c$_p$(T) shown in sec.~\ref{sec:cp}), revealing a high frustration ratio $f = \frac{\Theta_{CW}}{T_{min}} > 500$. Here, $T_{min}$ is the lowest temperature of 0.1 K in our specific heat measurement. Overall, the magnetic susceptibility of BCNO is qualitatively similar to that of $\rm SrCu_{1/3}Nb_{2/3}O_3$ (SCNO), $\rm SrCu_{1/3}Ta_{2/3}O_3$ (SCTO), and $\rm BaCu_{1/3}Ta_{2/3}O_3$ (BCTO) studied previously~\cite{hossain2024evidence,sana2024possible,mahapatra2026emergent}.\\

 Furthermore, $\chi$ versus temperature plotted on the log-log scale in the insets of  Fig.~\ref{fig:chi_BCNO}(b), show a power-law (T$^{-\gamma}$) divergence below 10 K with exponent $\gamma \approx 0.6$, as expected for a random-singlet like phase. Unlike BCTO~\cite{mahapatra2026emergent}, where small deviation from the $T^{-\gamma}$ behavior begin to appear below about $4-5$~K, in BCNO the $T^{-\gamma}$ behavior remains down to the lowest measurement temperature. A similar type of  behavior was also reported previously in double perovskite based triangular lattice compound $\rm Y_2CuTiO_6$ due to random site-mixing of $\rm Cu^{2+}$ and $\rm Ti^{4+}$, leading to the formation of random-singlets at low temperatures~\cite{18_SKundu_YCTO}. The same power-law divergence in $\chi$(T) with nearly the same value of the exponent $\gamma$ is also observed in its sister compounds SCNO, SCTO where a random singlet phase has been reported due to strong disorder between Cu and the non-magnetic Nb/Ta~\cite{hossain2024evidence,sana2024possible}. %While an emergent random singlet state happens in BCTO, however the power-law divergence deviates below 4-5~K, eventually transforming in to a gapless spin liquid like phase~\cite{mahapatra2026emergent}.  
 \\

As another signature of a random-singlet phase, Sana et al.~\cite{sana2024possible} showed the scaling of susceptibility, $\rm (\mu_0H)^{\gamma}\chi~\propto~ T/\mu_0H$, under various magnetic fields in SCTO, where all the data points collapse on a single curve. In BCNO a similar data collapse can be seen, as shown in Fig.~\ref{fig:chi_BCNO}(c) using $\gamma~\approx~0.6$. Similarly, the isothermal magnetization (M) curves, measured at various temperatures, also collapse on a single curve when $\rm MT^{\gamma-1}$ is plotted against $\mu_0H$/T for ${\gamma-1} \approx -0.4$, as shown in Fig.~\ref{fig:chi_BCNO}(d). These observations supports the formation of a random-singlet phase in BCNO at low temperatures.\\

\subsubsection{Specific Heat}
\label{sec:cp}

\begin{figure*}
   \centering
   \includegraphics[width= \linewidth]{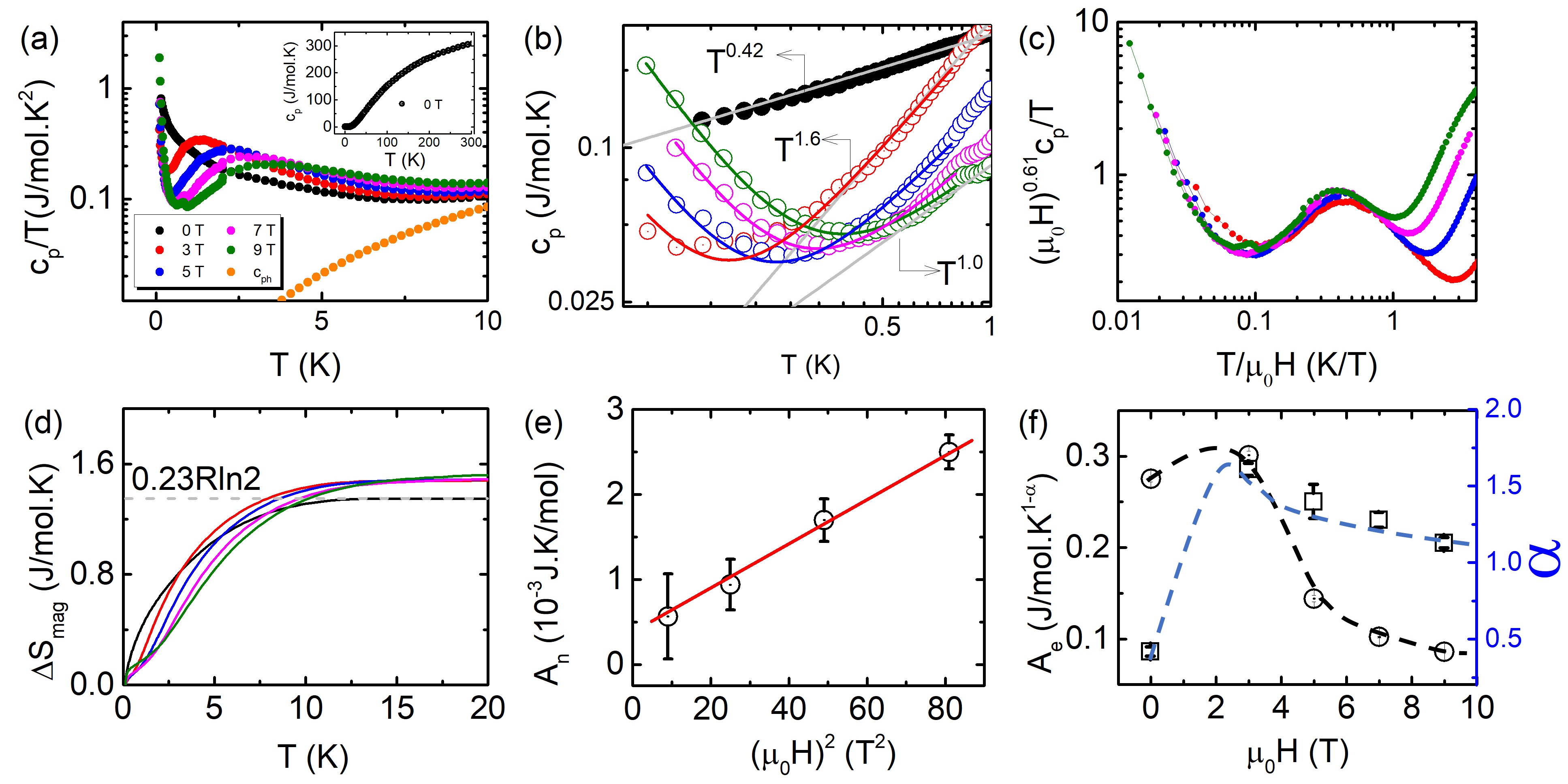}
  \caption{(a) The temperature variation of specific heat plotted
as $c_p/T$. The calculated phonon contribution is also shown. Inset: The temperature-dependent zero-field specific heat $c_p(T)$ in the temperature range of 0.1 $\le$ T $\le$ 300~K. (b) $c_p$ versus $T$ for various applied magnetic fields plotted on a log-log scale. The fitted curves (eq.~\ref{cplt}) are shown as solid lines through the data points. The gray colored lines are a guide to eye showing $T^{0.42}$ (representative of the zero-field data), $T^{1.6}$ (representative of the intermediate fields), and $T^{1.0}$ (representative of the high-field data). (c) The data collapse of the scaled specific heat $\rm \mu_0H^\gamma c_p/T$ plotted against the normalized
temperature T/$\mu_0$H for various applied fields. (d) The entropy change $\Delta S_{mag}$ above $0.1$~K under various magnetic fields. (e) The plot of the coefficient $\rm A_n$  of the nuclear Schottky term in eq.~\ref{cplt} plotted as a function of the square of the applied magnetic field. The solid red line is a linear fit to the data. (f) The magnetic field dependence of the coefficient $A_e$ (left $y$-axis) and exponent $\alpha$ (left $y$-axis) of the second term in eq.~\ref{cplt}, respectively.  } 
  \label{fig:BCNO-cp}
\end{figure*}

The low-temperature specific heat ($c_p$) of BCNO is shown in Fig.~\ref{fig:BCNO-cp}(a), where it is plotted as $c_p/T$ %in the temperature range 0.1 $\le$ T $\le$ 300~K 
under various applied magnetic fields ($\mu_0$H = 0, 3, 5, 7, 9~T). Upon cooling in zero-field, $c_p/T$ decrease smoothly down to about 8~K and thereafter shows an upturn leading to a diverging behavior without any signs of long-range magnetic ordering down to 0.1~K. In the presence of a magnetic field, the upturn shifts to higher temperatures, leading to a broad peak at lower temperatures. The peak position and peak width are sensitive to the applied field: as the field increases, the peak broadens unusually and shifts to higher temperatures. This unusual $c_p[H, T]$ dependence cannot be explained using a simple two-level Schottky-type scenario with a fixed energy gap. This is due to the fact that in BCNO (and other such systems characterized by the formation of random singlets) this anomaly results from an exchange-induced splitting $J$ that has a broad and continuous distribution. Its evolution under magnetic field is , therefore, more complex, emerging from the interplay of $J$ and the Zeeman energy~\cite{kimchi_nat_com_scaling}. \\ 

The magnetic contribution $ c_{mag}(T, H)$, shown in Fig.~\ref{fig:BCNO-cp}(b), is derived by subtracting $c_{ph}$ from the $c_p$ data. For this, $c_p$ is fitted between 15 K and 25 K using the equation:
\begin{equation}
\label{eq:cph}
c_p (T) = \beta T^3 + \delta T^5 + \theta T^7,    
\end{equation}
 where T$^5$ and T$^7$ terms are included to improve the quality of the fit. The higher order terms generally comes from lattice anharmonicity, but in the present case, since the Curie-Weiss temperature is high (higher than the fitting range), the departure from $T^3$ may also arise from small magnetic contribution. However, as we shall see below, inclusion or non-inclusion of these terms has hardly any consequence on our interpretation of the ground state.\\
 
The values of the coefficients for the best-fit came out to be: $\rm \beta = 7.647\times10^{-4}~J~mol^{-1} K^{-5}$, $\rm \delta = 5.408\times10^{-8}~J~mol^{-1} K^{-6}$, and $\rm \theta = 5.707\times10^{-11}~J~mol^{-1} K^{-8}$. As expected, the values of $\delta$ and $\theta$ are three to four orders of magnitude smaller than the coefficient of the cubic term. A more detailed discussion on the inclusion/non-inclusion of these terms is provided in the Supplementary Material~\cite{Suppl_mat}, where we also discussed the effect of fitting temperature range on the extracted $c_{mag}$. Details apart, the main conclusion that we arrived at is that $c_{mag}$ below 10~K, where the contribution from these terms become extremely small, is pretty rugged independent of whether we use these terms or not.\\ 

Furthermore, below about 2 K, the phonon contribution, relative to the magnetic contribution, itself become too weak to be of any practical consideration (See Fig. 5a where $c_{ph}$ is also plotted). Thus, below 2 K, one can as well approximate the magnetic specific heat, $c_{mag}$, by the total measured specific heat $c_p$, as is generally done in the literature. However, using eq.~\ref{eq:cph}, a reasonable estimate of $c_{mag}$ at high temperatures can nevertheless be obtained using the $\beta$ values listed above, which will be useful in estimating the magnetic entropy. Depending upon whether the higher-order terms are included or not in estimating $c_{ph}$, the estimated entropy can vary within $\pm~10\%$. \\

c$_{mag}[H, T]$, Fig. \ref{fig:BCNO-cp}(b), under zero magnetic field shows a power-law dependence, characteristic of the random-singlet phase, with $c_{mag}~\propto~T^{1-\gamma}~=~T^{0.42}$ up to temperatures as high as 3~K, i.e., over more than a decade in temperature (The $T^-{0.42}$ fit up to 3 K is shown in Supplementary Material~\cite{Suppl_mat}). Furthermore, the value of $\gamma$ obtained from exponent $0.42$ is also in excellent agreement with that obtained from the diverging behavior of $\chi(T)$ and $M[H, T]$-scaling. Note that same behavior can be shown for $c_p$ as well, which is essentially equal to $c_{mag}$ as $c_{ph} \ll c_{mag}$ over this temperature range (see Fig.~\ref{fig:BCNO-cp}(a), where $c_{ph}$ derived using eq.~\ref{eq:cph} is shown).\\  
\par When a magnetic field is applied, $c_{mag}$, which in zero-field follows a $T^{0.42}$ dependence down to the lowest temperature, shows an upturn. Furthermore, this upturn scales with the field, becoming stronger as the field strength increases, which suggests that the upturn possibly originates from the nuclear hyperfine splitting. The other notable observation relates to the dramatic restructuring of the power-law temperature dependence of the specific heat, which changes from $T^{0.42}$ in zero-field to almost linear in $T$ as $\mu_0H \rightarrow 9$~T, with an intermediate region where the exponent first increase before decreasing to a value of $\approx 1.1$ at 9 T, indicating a field-driven crossover from a disordered random singlet phase to a $T$-linear regime, which, in frustrated magnetic insulators, generally signify emergence of gapless magnetic excitations. Without committing ourselves to any interpretation, let us examine these possibilities further in greater detail.\\
\par To systematically track this evolution, we fitted the low-temperature ($T \lesssim 1\text{ K}$) data using the expression: \begin{equation} \label{cplt}
c_p = \frac{A_n}{T^2} + A_eT^{\alpha},
\end{equation} where the first term is to account for the nuclear Schottky contribution ($A_n$) wherein $1/T^2$ variation approximates the high-temperature tail of the nuclear Schottky anomaly; and, the second term captures the magnetic specific heat ($A_e T^{\alpha}$). The exponent $\alpha$ relates to the scaling parameter through the relation: $\alpha = 1-\gamma$. The field dependence of these parameters is displayed in Fig.~\ref{fig:BCNO-cp}(e, f), where $A_n$ is plotted against $H^2$ in panel (e), and $A_e$ and $\alpha$ in panel (f).\\
\par We note that the coefficient $A_n$ varies linearly with $H^2$, confirming the Schottky-like origin of the first term in eq.~\ref{cplt}. At low fields ($H \rightarrow 0$), the system is dominated by the random singlet physics. In this regime, the nuclear contribution $A_n$ is negligible as the nuclear levels remain nearly degenerate. The RS phase is therefore clearly manifested in the zero-field behavior of $c_{mag}$ that varies as $T^\alpha$ ($\alpha = 0.42$), a direct signature of the sub-linear power-law scaling expected for a random distribution of exchange-coupled dimers. The exponent value is also in close agreement with scaling parameter obtained from magnetization data as mentioned earlier.\\
As the magnetic field increases, somewhere between $3$ to $4$~T, one sees a maximum in the value of $\alpha$ and $A_e$. Above the maximum, both $\alpha$ and $A_e$ exhibit a decreasing behavior, and as $H \rightarrow 9$~T, the $A_e$ versus $H$ plot tends to flatten with $A_e$ approaching a value close to $0.075$~J mol$^{-1}$ K$^{-2}$ ($\sim 75$~mJ mol$^{-1}$ K$^{-2}$). At the same time, the exponent $\alpha$ converges toward $1$. What this possibly indicate is that the intrinsic magnetic specific heat of BCNO tends to approach a linear-in-temperature behavior under high magnetic fields, suggestive of the emergence of a quantum paramagnetic regime, possibly a quantum spin liquid. Notably, the coefficient of the linear term, $\sim 75$~mJ mol$^{-1}$ K$^{-2}$, is comparable to a similar value reported in literature for several gapless quantum spin liquids (see Table II in Supplementary Material~\cite{Suppl_mat}). The observed crossover also seems to agree with the proposal by Kimchi~\textit{et al.}, who argued that quantum systems with quenched disorder can be described in terms of two interrelated subsystems: ``a fraction of spins form random valence bonds $\cdots$~surrounded by a quantum paramagnetic phase''~\cite{kimchi_nat_com_scaling}. Therefore, it is possible that in BCNO, the random singlets dominate the thermodynamic response under zero-field, and under high fields the quantum paramagnetic phase reveals its presence in $c_p$; however, more complementary experiments should be carried out to confirm this proposition.\\
\par To estimate the fraction of Cu spins forming random valence bonds (i.e., random-singlet phase) down to 0.1 K, we calculate the magnetic entropy. If all the spins in the system nucleate to form spin-singlets, the magnetic entropy should approach $R\ln2 = 5.76$~J mol$^{-1}$ K$^{-1}$, signifying the absence of any residual magnetic entropy. So, the estimation of magnetic will lend further insight about the nature of the ground state in BCNO. %However, as argued above, only a fraction of spins participate in the random-singlet phase, while the larger fraction remains un-nucleated or dynamic (the quantum paramagnetic phase), and therefore we expect a considerable residual spin entropy in BCNO.\\ 

\par The magnetic entropy, $\Delta$S$_{mag}$, is obtained by integrating $ c_{mag}/T$ above $0.1$~K for various magnetic fields as shown in Fig.~\ref{fig:BCNO-cp} (d). In zero-field, $\Delta S_{mag}$ saturates around 1.34~J mol$^{-1}$ K$^{-1}$ above a temperature of $\approx10$ K. The saturated entropy is about 23\% of $R\ln2$, suggesting that approximately one-fourth of the spins contribute to the low-energy random-singlet phase down to 0.1 K%, leaving about three-fourths of the spins in the quantum paramagnetic phase, at least down to 0.1 K
. It is possible that the remaining three-fourth of the spins are very weakly coupled ($J \lesssim 0.1$~K) and therefore have not nucleated, or it might also be that they represent a quantum paramagnetic phase, which is expected in this phenomenology, at least in the context of geometrically frustrated system~\cite{kimchi_2018, kimchi_nat_com_scaling}, and is supported by the linear-in-T $c_p$. Similarly high values of residual entropy has been previously reported for a number of spin-1/2 systems with coexisting random singlet and spin liquid phenomenology~\cite{hossain2024evidence,sana2024possible,mahapatra2026emergent,7_BCSO_Zhou,8,18_SKundu_YCTO}. Further experiments are needed to clarify this.\\ 
\par In Fig.~\ref{fig:BCNO-cp}d, it is evident that Under applied magnetic fields recovered entropy above 0.1 K saturates at a slightly higher value. What contributes this additional spin entropy? One can envisage two plausible additional factors that may contribute to $\Delta S_{mag}$: (i) the `orphan' spins, i.e., the Cu atoms not occupying their regular B-site; for example, located at the interstitial sites in the lattice, and (ii) the nuclear Schottky contribution ($A_n/T^2$ term in eq.~\ref{eq:cph}). Let us estimate the nuclear Schottky contribution first. This contribution above 0.1 K will be higher, the higher the applied field. We thus make an estimate on its upper limit by calculating the $\Delta S_{nuclear}$ by integrating $A_n/T^2$  under 9 T from 0.1 K to 20 K using the value of $A_n$ from the low temperature fit (eq.~\ref{eq:cph}). This gives $\Delta S_{nuclear} \approx 0.12$~J mol$^{-1}$ K$^{-1}$, which saturates at this value within about 1-2 K of the temperature rise above 0.1 K. The same calculation for the 3 T field yields a value of $\approx0.04$~J mol$^{-1}$ K$^{-1}$. \\

However, the observed increase in $\Delta S_{mag}$ under 3 T, i.e., $\Delta S_{mag} (\mu_0H = 3 T) - \Delta S_{mag} (\mu_0H = 0)$ is close to 0.14 mol$^{-1}$ K$^{-1}$, implying that the additional $0.10$~J mol$^{-1}$ K$^{-1}$ comes from the orphan spins. Similarly, the observed increase in $\Delta S_{mag}$ under 9 T is approximately $1.52 - 1.34 = 0.18$~J mol$^{-1}$ K$^{-1}$. Since, under 9 T, $\Delta S_{nuclear} \approx 0.12$~J mol$^{-1}$ K$^{-1}$, the contribution of the orphan spins can be estimated to be $\approx 0.18 - 0.12 = 0.06$~J mol$^{-1}$ K$^{-1}$. Though not exactly the same, this estimate is in the same ballpark as the one obtained using the 3 T data. The absolute difference between the two is well within the collective uncertainties in the measurement of $c_p$ and various approximations involved in estimating these minor contributions.\\

We can now estimate the concentration of the orphan spins. To set an upper limit on this concentration, we take the higher of the two values above (0.1~J mol$^{-1}$ K$^{-1}$ under 3 T). This in turn gives $\frac{0.1}{5.76}\times 100 \approx 1.7\%$ as the upper limit on the fraction of the orphan spins. This estimate looks fairly reasonable, given that in zero-field we barely see any Schottky-like bump or an upturn down to very low temperatures, as is generally seen when the orphan spin concentration is 3-4\% or higher~[see Supplementary Material in Ref. \cite{hossain2024evidence}].\\ 

%The high residual entropy ($\approx 75$\% of $R\ln2$) can be attributed to the highly degenerate quantum paramagnetic phase (the second subsystem in Ref.~\cite{kimchi_nat_com_scaling}).      

\subsection{Discussion}
We now discuss the possible random spin singlet phenomenology in BCNO and its evolution in presence of large applied magnetic field which helps reveal the quantum paramagnetic phase.\\

The exponent $\gamma \approx 0.6$, obtained from the power-law behavior of $\chi(T)$ and $M[H, T]$-scaling, also fits the low temperature specific heat satisfactorily which not only shows a $T^{1-\gamma} \approx T^{0.42}$ dependence over a wide temperature range, but also the single parameter data scaling behavior over a wide range of temperature and magnetic fields, as shown in Fig.~\ref{fig:BCNO-cp}c. These observations solidly establish the ground state of BCNO in terms of a random-singlet phase.\\ 

The question one might ask is: since the concentration of Cu (33\%) is almost as large as the percolation threshold of a cubic lattice (31\%), in BCNO %where Cu and Nb are distributed over a pseudo-cubic lattice in 1:2 ratio, should 
one expects Cu atoms to form large, connected, chains or clusters, and such clusters should have favored a glassy/Griffiths-like phase if not a long-range ordered ground state. Then why do we see a random singlet phase and not a more conventional and conceivable behavior. Our EXAFS analysis helped resolve this puzzle. We show that the distribution of Cu and Nb in the lattice depicts high hetero-atomic correlation maximizing the Cu monomers.\\

Finally, the application of external magnetic field led to a linear-in-T specific heat suggestive of the presence of a quantum paramagnetic phase. Further complementary experiments including NMR, $\mu SR$ and inelastic neutron scattering under high field should be done to confirm if the T-linear behavior is indeed to a quantum paramagnetic phase or not.\\

It is also instructive to compare the magnetic ground-state properties of BCNO with those reported for the related disordered perovskites SCNO, SCTO, and BCTO. Among these compounds, SCNO exhibits the closest phenomenology to BCNO with satisfactory $T/H$ scaling and power-law behavior of $\chi(T)$ and $c_{mag}(T)$ expected for a random-singlet state. Furthermore, under applied magnetic field, the low-temperature specific heat evolves toward an approximately linear-in-(T) form with a coefficient of nearly 50 mJ~mol$^{-1}$~K$^{-1}$ (see Fig. S3 in Ref.~\cite{hossain2024evidence}). \\

In contrast, BCTO exhibits noticeably different low-temperature behavior. Although the susceptibility follows an approximate power-law dependence over an intermediate temperature range, deviations appear below about 5 K, and the characteristic (T/H) scaling associated with the random-singlet phase is absent. Moreover, the zero-field specific heat already displays a nearly linear temperature dependence at low temperatures. These differences suggest that, despite the close structural relationship among these compounds, their low-energy magnetic excitations may be governed by distinct distributions of exchange couplings.\\ 

SCTO appears to occupy an intermediate position. Power-law behavior is observed in both $\chi(T)$ and $c_{mag}$; however, the exponent $\gamma$ from $\chi(T)$ ($=~0.57$) differs from $c_{mag}$ ($= 0.41$. Similar discrepancies between exponents extracted from different thermodynamic probes have also been reported in other random-singlet candidates and remain an open issue. Taken together, these observations suggest that while random-singlet phenomenology may be a recurring feature of disordered Cu-based perovskites, the detailed low-energy behavior is highly sensitive to the underlying disorder and exchange-network connectivity. Further studies will throw light on these aspects of the random-singlet physics in higher dimensions.  

\section{Summary and conclusions} 
We investigated the disordered perovskite BaCu$_{1/3}$Nb$_{2/3}$O$_3$ (BCNO). Its structure was characterized using high-resolution synchrotron X-ray powder diffraction and XAFS spectroscopy. Rietveld refinement of the XRD data shows that BCNO adopts a pseudo-cubic structure, i.e., a tetragonally distorted variant of the ideal cubic perovskite, which is well described by the $P4mm$ space group with c/a $\approx$ 1.036. The XAFS analysis confirms that Cu and Nb ions are locally distributed in a uniform $1:2$ ratio, but with strong hetero-atomic correlation at the local scale favoring the formation of Cu monomers. This rules out the presence of magnetic clusters that might otherwise give rise to a spin-glass-like ground state. Notably, the analysis indicates that, on average, the Nb absorber has 3 Cu and 3 Nb ions as neighbors, whereas a Cu absorber is coordinated by fewer than 0.6 Cu and more than 5.4 Nb ions. This finding is significant and is supported by the probabilistic arguments illustrated in Fig.~1 of the Supplementary Material (see Ref.~\cite{mahapatra2026emergent}). These results indicate that the dominant magnetic exchange pathway between two Cu ions is Cu-O-Nb-O-Cu, with very rare contribution from direct Cu-O-Cu connections. Furthermore, the number of Nb ions participating in the superexchange path cannot increase without bound as each Nb finds around it as many Cu as Nb, ruling out the formation of long Nb chain between two Cu.\\

\par The magnetization measurements exhibit no sign of long-range magnetic ordering or spin-glass-like freezing (due to the absence of magnetic clusters as inferred from EXAFS), rather the magnetic susceptibility follows an approximate power-law behavior above 3 K, $\chi \propto T^{-\gamma},~\gamma~\approx~0.6$ and a single parameter data collapse in the $(\mu_0H)^{0.6}\chi$ vs. T/$\mu_0H$ plot. Additionally, the $M(H, T) $-scaling works reasonably well, as demonstrated by $\rm MT^{0.4}$ versus $\rm \mu_0H/T$ plots, which show the necessary data collapse. Furthermore, the expected power-law behavior for $\rm c_{mag}$ ($\rm c_{mag} \propto T^{1-\gamma} \approx T^{0.4}$) could be detected in the 0.1-3~K range. Additionally, $c[H, T]$-scaling, i.e., $(\mu_0H)^{\gamma}c_p/T$ versus T/$\mu_0$H plots show the requisite data collapse for $\gamma \approx 0.6$ (the power-law exponent in $\chi$(T)). 
These observations confirm the formation of a random singlet like phase in BCNO. However, the entropic consideration suggests that only about 25\% of the spins are involved, the 75\% remaining seems to constitute the background quantum paramagnetic phase. Our work should motivate further studies to confirm the presence of the quantum paramagnetic phase.  
Establishing unambiguously the presence of both subsystems with two distinct phases: random singlet phase and quantum paramagnetic phase, in BCNO will be useful in establishing the general properties of the random singlet phase in three-dimensional systems.

\begin{acknowledgments}
Portions of this research were carried out at the light source PETRA-III of DESY, a member of the Helmholtz Association (HGF). We would like to thank the beamline scientists for assistance at the beamlines P65 and P02.1. Financial support by the Department of Science \& Technology (Government of India) provided within the framework of the India@DESY collaboration is gratefully acknowledged. 
SM would like to acknowledge the University Grants Commission (UGC), India, for financial support in the form of a research fellowship. SM is also grateful to the I-HUB Quantum Technology Foundation (QTF) at IISER, Pune, for financial support under the Senior Research Fellowship scheme. SM acknowledges Dr V.S. Patankar Dissertation Fellowship for supporting through a one-time financial aid. We thank Pramod Nadig for help with the magnetic measurements.  

\end{acknowledgments}

\bibliography{mybib.bib}
\bibliographystyle{apsrev4-2}

\end{document}